\documentstyle[11pt]{article}
\def\bkR{{\rm I\kern-.17em R}}
\def\bkC{{\rm \kern.24em \vrule width.05em height1.4ex depth-.05ex \kern-.26em C}}

\newcommand{\R}{{\sf R\hspace*{-0.9ex}\rule{0.15ex}%
{1.5ex}\hspace*{0.9ex}}}
\def\bkR{{\rm I\kern-.17em R}}
\def\bkC{{\rm \kern.24em \vrule width.05em height1.4ex depth-.05ex \kern-.26em C}}

\begin{document}

\author{Nuno Costa Dias\footnote{{\it ncdias@mail.telepac.pt}} \\ Jo\~{a}o Nuno Prata\footnote{{\it joao.prata@ulusofona.pt}} \\ {\it Departamento de Matem\'atica} \\
{\it Universidade Lus\'ofona de Humanidades e Tecnologias} \\ {\it Av. Campo Grande, 376, 1749-024 Lisboa, Portugal}\\
{\it and}\\
{\it Grupo de F\'{\i}sica Matem\'atica}\\
{\it Universidade de Lisboa}\\
{\it Av. Prof. Gama Pinto 2}\\
{\it 1649-003 Lisboa, Potugal}}

\title{Features of Moyal Trajectories}

\maketitle
\begin{abstract}
We study the Moyal evolution of the canonical position and momentum variables. We compare it with the classical evolution and show that, contrary to what is commonly found in the literature, the two dynamics do not coincide. We prove that this divergence is quite general by studying Hamiltonians of the form $p^2 /2m + V(q)$. Several alternative formulations of Moyal dynamics are then suggested. We introduce the concept of starfunction and use it to reformulate the Moyal equations in terms of a system of ordinary differential equations on the noncommutative Moyal plane. We then use this formulation to study the semiclassical expansion of Moyal trajectories, which is cast in terms of a (order by order in $\hbar$) recursive hierarchy of i) first order partial differential equations as well as ii) systems of first order ordinary differential equations. The latter formulation is derived independently for analytic Hamiltonians as well as for the more general case of smooth local integrable ones. We present various examples illustrating these results. 
\end{abstract}

\section{Introduction}

In recent years there has been a lot of work devoted to noncommutative structures in mathematics and physics \cite{Nekrasov}, \cite{Witten}. In most cases one deals with the Moyal product, which first appeared in the context of deformation quantization \cite{Weyl,Wigner,Groenewold,Moyal}. If we consider the position and momentum operators generators of the Heisenberg algebra ${\cal A}$, then we can define the associated universal enveloping algebra in quantum phase space ${\cal U}_{\star} \left( {\cal A} \right)$, where c-functions are multiplied by resorting to a twisted convolution which admits a formal $\hbar$-expansion\footnote{We shall only consider one-dimensional systems in this work. The generalization to higher dimensions is straightforward.}:
\begin{equation}
f(q,p) \star g(q,p) = \left. \exp \left[\frac{i \hbar}{2} \left( \frac{\partial}{\partial q}\frac{\partial}{\partial p'}-\frac{\partial}{\partial p}\frac{\partial}{\partial q'} \right) \right] f(q,p) g(q',p') \right|_{(q',p') = (q,p)},
\end{equation}
or an alternative kernel representation valid in the set of phase space square integrable functions $L_2(\bkC,d\Gamma)$, $d\Gamma=dqdp$ \cite{Bracken1}:
\begin{equation}
f(q,p) \star g(q,p) = \frac{1}{( \pi \hbar )^2} \int d q' d p' dq'' dp'' \hspace{0.2 cm} f(q',p') g(q'' ,p'') \exp \left\{- \frac{2i}{\hbar} \left[p(q'-q'') + p'(q''-q) +p''(q-q') \right] \right\}.
\end{equation}
This is the so-called Moyal (or Groenewold) product. One can also construct a Lie algebraic bracket - the Moyal bracket - according to:
\begin{equation}
\left[f(q,p) , g(q,p) \right] = \frac{2}{\hbar} \left. \sin \left[\frac{\hbar}{2} \left( \frac{\partial}{\partial q}\frac{\partial}{\partial p'}-\frac{\partial}{\partial p}\frac{\partial}{\partial q'} \right) \right] f(q,p) g(q',p') \right|_{(q',p') = (q,p)}.
\end{equation}
These provides the basic structure for the formulation of quantum mechanics in the phase space \cite{Bracken1}-\cite{Balazs}. One of the remarkable properties of these expressions is the fact that they constitute formal deformations of the usual commutative product and of the Poisson bracket, respectively:
\begin{equation}
\left\{
\begin{array}{l}
f(q,p) \star g (q,p) = f(q,p) \cdot g (q,p) + {\cal O} (\hbar)\\
\\
\left[f(q,p) , g (q,p) \right] = \left\{ f(q,p) , g (q,p) \right\} + {\cal O} (\hbar^2)
\end{array}
\right.
\end{equation}
The quantum phase space endowed with the Moyal bracket admits a Heisenberg evolution for (real) c-functions according to:
\begin{equation}
\dot A (q,p,t) = \left[A(q,p,t) , H (q,p) \right]_{(q,p)}, \hspace{1 cm} A\left(q,p, t=t_0 \right) =A_0 (q,p)
\end{equation}
where $H(q,p)$ is the Hamiltonian symbol (taken to be real and $\hbar$-independent). In particular, for the fundamental variables $(q,p)$, we have:
\begin{equation}
\dot Q (q,p,t) = \left[Q(q,p,t) , H (q,p) \right]_{(q,p)}, \hspace{0.5 cm} \dot P (q,p,t) = \left[P(q,p,t) , H (q,p) \right]_{(q,p)},
\end{equation}
subject to the initial conditions:
\begin{equation}
Q(q,p,t=t_0) = q , \hspace{1 cm} P (q,p,t=t_0) =p.
\end{equation}
The subscript $(q,p)$ in eq.(6) stresses the fact that the Moyal bracket is to be evaluated with respect to the canonical set $(q,p)$. From (6,7) we may then write at time $t=t_0$:
\begin{equation}
\dot Q (q,p, t=t_0) = \frac{\partial}{\partial p} H (q,p), \hspace{0.5 cm} \dot P (q,p, t=t_0) = - \frac{\partial}{\partial q} H (q,p).
\end{equation}
Equations (8) are reminiscent of the classical Hamilton equations:
\begin{equation}
\dot Q (q,p, t) = \frac{\partial}{\partial P} H (Q,P), \hspace{0.5 cm} \dot P (q,p, t) = - \frac{\partial}{\partial Q} H (Q,P).
\end{equation}
It is commonly claimed that, because of eqs.(8), the fundamental $(q,p)$ evolve classically. The difference between Moyal and classical evolution would then only become apparent once functions of q's and p's are considered.

The subject of Moyal dynamics is a subtle one and previous work in the field is scarce. In \cite{Flato} the difference between classical and Moyal dynamics for general observables is clearly acknowledged, although no concrete statement is made concerning the canonical variables. In \cite{Dias4} we pointed out explicitly that, in the Moyal framework, Hamilton's equations for the fundamental variables (8) should be solved in the noncommutative algebra ${\cal U}_{\star} \left({\cal A} \right)$ (in which case they become valid at later times). The formulation of Weyl-Wigner quantum mechanics in the Heisenberg picture (a key ingredient for the discussion of Moyal trajectories) has been studied in many papers (see for instance \cite{Dias2,Osborn}). One of the most significative papers in the field of Moyal dynamics is \cite{Osborn}, where the Moyal correction to the classical Hamiltonian equations is derived order by order in $\hbar$ by using a cluster-graph representation of the star-exponential as well as a recursive hierarchy of classical transport equations (see also \cite{Prosser,Berezin}). We shall review this method in section 3.1.  
 
The purpose of this paper is to further clarify certain aspects of Moyal dynamics and present some new results. In particular we shall prove with concrete examples that the claim about the coincidence of Moyal and classical trajectories is false for most Hamiltonians. The subject of trajectories generated by a bracket on a symplectic (or more generally a Poisson manifold) with a star product is a notoriously difficult one. The present work is devoted to the simpler case of the flat phase space of a one dimensional system $T^*M \simeq \R^2$ with a global Darboux chart and symplectic form $\omega = dq \wedge dp$. We shall elaborate on the specific features of Moyal trajectories which make them distinct from their classical (Hamilton) counterparts, namely the facts that, in general, Moyal evolution does not act as a coordinate transformation nor as a symplectomorphism in phase space. Moreover we propose strategies for solving the equations, if not completely then at least formally or semiclassically (in a $\hbar$ power series). These results are developed using noncommutative phase space methods while never resorting to operator methods.

We feel that Moyal dynamics is an interesting and challenging mathematical problem {\it per se}. However Moyal trajectories are also expected to play an important role in certain physical situations. In ref.\cite{Mikovic} we proved that the expectation value of an observable $\hat A$ (typically the position operator) in a state with density matrix $\hat{\rho}$ is given by:
\begin{equation}
A(t) \equiv Tr \left( \hat A (t) \hat{\rho} \right) = \left. \tilde F \left( - i \frac{\partial}{\partial z} \right) A_M (z,t) \right|_{z=0}.
\end{equation}
In the previous equation $z=(q,p)$, $\frac{\partial}{\partial z}  = \left(\frac{\partial}{\partial q}, \frac{\partial}{\partial p} \right)$, $A_M(z,t)$ is the Moyal  evolution of $A(z)$ (in other words, the Weyl symbol of the Heisenberg operator $\hat A (t)$ ) which is the solution of eq.(5) and $\tilde F (a)$ is the symplectic Fourier transform (or chord function) of the Wigner function $F(z)$:
\begin{equation}
\tilde F (a) = \int dz \hspace{0.2 cm} F(z) e^{i \sigma (a,z) }, \hspace{1 cm} \sigma (a,z) = a_q p -a_p q.
\end{equation}  
If the state of the system is described by a coherent state with wavefunction
\begin{equation}
\psi_{\alpha} (x) = \left( \frac{m \omega}{\pi \hbar} \right)^{\frac{1}{4}} \exp \left[ - \frac{m \omega}{2 \hbar} \left(x-q \right)^2 + \frac{i p}{\hbar} \left(x- \frac{q}{2} \right) \right], \hspace{1 cm} \alpha = \sqrt{\frac{m \omega}{2 \hbar}} q + \frac{i p}{\sqrt{2 m \omega \hbar}},
\end{equation}
then eq.(10) reduces to:
\begin{equation}
A(t) = \exp \left[ \frac{\hbar}{4 m \omega} \frac{\partial^2}{\partial q^2} +  \frac{\hbar m \omega}{4} \frac{\partial^2}{\partial p^2} \right] A_M (q,p,t).
\end{equation}
This expression is particularly well suited for the semiclassical expansions in powers of Planck's constant. It is well known that expectation values such as (13) (or more generally (10)) play an important role in quantum optics \cite{Lee} and in determining effective actions in quantum mechanics and quantum field theories \cite{Jona}, \cite{Zappala}. Moreover the method of trajectories finds important applications in quantum hydrodynamics \cite{Wyatt} and transport models in quantum chemistry and heavy-ion collisions \cite{Aichelin} and it is also expected to be crucial if one aims at fully grasping the nonlocal nature of noncommutative quantum field theories \cite{Nekrasov}. 

We shall start by recapitulating some well known features of classical dynamics, namely that classical evolution is a canonical and a coordinate transformation, and that the Hamiltonian is a constant of motion when not explicitly time dependent \cite{Goldstein}, \cite{Arnold} (section 2). We shall then show how these facts translate to Moyal dynamics (section 2) and propose new formulations of Moyal equations and new strategies to solve them (section 3). All the results will be illustrated with examples (section 4). The Hamiltonian of the form $H = \frac{p^2}{2m} + V(q)$ will be studied in detail (section 5).

\section{Classical and Moyal dynamics}

Some of the results of this section are well known. We present them here for completeness and to highlight the differences between Moyal and classical evolutions. We will present simple proofs of the results whenever we feel that there is some subtlety that should be emphasized. We shall use the subscripts $C$ and $M$ for classical and Moyal evolutions, respectively. 

One possible formulation of classical dynamics is in terms of a classical transport equation. Let then $A_C (q,p,t)$ be the solution of the first order partial differential equation
\begin{equation}
\dot A_C (q,p,t) = \left\{A_C (q,p,t), H (q,p) \right\}_{(q,p)},
\end{equation}
subject to the initial condition
\begin{equation}
A_C (q,p, t=t_0) = A_0 (q,p).
\end{equation}

\vspace{0.3 cm}
\noindent 
Then it is easy to prove that:

\vspace{0.3 cm}
\noindent
{\underline{\bf Theorem 2.1:}} The solution is
\begin{equation}
A_C (q,p,t) = A_0 \left( Q_C (q,p,t), P_C (q,p,t) \right),
\end{equation}
where $\left(Q_C (q,p,t), P_C (q,p,t) \right)$ are the solutions of Hamilton's equations (9) with initial conditions (7).

\noindent
By resorting to Jacobi's identity and theorem 2.1, we can also show that:

\vspace{0.3 cm}
\noindent
{\underline{\bf Theorem 2.2:}} The transformation 
\begin{equation}
(q,p) \longrightarrow \left(Q=Q_C (q,p,t) , P = P_C (q,p,t) \right)
\end{equation}
is canonical, i.e.
\begin{eqnarray}
& \left\{A \left(Q_C (q,p,t), P_C (q,p,t) \right) , B \left(Q_C (q,p,t), P_C (q,p,t) \right) \right\}_{(q,p)} & \nonumber \\
&= \left\{A(Q_C(q,p,t),P_C(q,p,t)) , B (Q_C(q,p,t),P_C(q,p,t)) \right\}_{(Q_C,P_C)},&
\end{eqnarray}
for general phase space functions $A,B$.

\vspace{0.3 cm}
\noindent
{\underline{\bf Lemma 2.3:}} The Hamiltonian is a constant of motion.

\vspace{0.3 cm}
\noindent
{\underline{\bf Proof:}} For an arbitrary observable we have:
\begin{equation}
\dot A_C (q,p,t) = \left\{A_C (q,p,t), H_0 (q,p) \right\}_{(q,p)} \equiv F (q,p,t),
\end{equation}
where we introduced the notation $H_0 (q,p) = H(q,p,t=t_0)$. Differentiating the previous equation with respect to time, we obtain:
\begin{equation}
\dot F (q,p,t) = \left\{\dot A_C, H_0 \right\} + \left\{A_C, \dot H_0 \right\} = \left\{F (q,p,t), H_0 (q,p) \right\}_{(q,p)}.
\end{equation}
From theorem 2.1 it then follows:
\begin{equation}
F(q,p,t) = F \left(Q_C (q,p,t), P_C (q,p,t), t=t_0 \right).
\end{equation}
In particular, let us choose $A=H$. Then, since $F(q,p,t=t_0)=0$, we conclude from the previous analysis that $F(q,p,t)=0$ and from (19):
\begin{equation}
\dot H (q,p,t) = \left\{H(q,p,t),H_0 (q,p)\right\}_{(q,p)} =0._{\Box} 
\end{equation}

\vspace{0.3 cm}
\noindent
An obvious corollary of the previous result is the following:

\vspace{0.3 cm}
\noindent
{\underline{\bf Corollary 2.4:}} The functional form of the Hamiltonian remains unchanged in the course of classical evolution:
\begin{equation}
H \left( Q_C (q,p,t), P_C (q,p,t) \right) = H (q,p).
\end{equation}

\noindent
The next theorem is the crux of the matter when we compare classical and Moyal evolutions. It states that the equations of motion are the same at any time $t$:

\vspace{0.3 cm}
\noindent
{\underline{\bf Theorem 2.5:}} The partial differential equations (14) for the position and momentum are equivalent to the set of ordinary differential equations:
\begin{equation}
\dot Q = \left\{Q,H \right\}_{(Q,P)} = \frac{\partial H}{\partial P}, \hspace{1 cm}  
\dot P = \left\{P,H \right\}_{(Q,P)} = - \frac{\partial H}{\partial Q}.
\end{equation}

\vspace{0.3 cm}
\noindent
{\underline{\bf Proof:}} We prove the statement for $Q$. From eq.(14), theorem 2.2, lemma 2.3 and its corollary we get:
\begin{equation}
\begin{array}{c}
\dot Q (q,p,t) = \left\{Q (q,p,t), H (q,p) \right\}_{(q,p)} = \left\{Q (q,p,t), H \left(Q(q,p,t) ,P (q,p,t) \right) \right\}_{(q,p)} =  \\
\\
=\left\{Q (q,p,t), H \left(Q(q,p,t) ,P (q,p,t) \right) \right\}_{(Q,P)} \Longleftrightarrow \dot Q = \left\{Q, H (Q,P) \right\}_{(Q,P)},
\end{array}
\end{equation}
where we omitted the subscript "{\it C}" for simplicity and used eq.(18).$_{\Box}$

\vspace{0.3 cm}
\noindent
{\underline{\bf Remark 2.6:}}
Notice that this presentation of classical dynamics is somewhat unusual. We posited the evolution to be given by the classical transport equation equation (14) even if $A_C$ stands for the position or the momentum. For instance:
\begin{equation}
\dot Q_C (q,p,t) = \left\{Q_C (q,p,t) , H (q,p) \right\}= \frac{\partial Q_C}{\partial q}\frac{\partial H}{\partial p} - \frac{\partial Q_C}{\partial p}\frac{\partial H}{\partial q}, \hspace{1 cm} Q_C (q,p,t= t_0) = q,
\end{equation}
and a similar equation for $P_C (q,p,t)$. This is a partial differential equation. It is only through theorem 2.5 that we show the coincidence of the solutions of the partial differential eq.(26) and of the system (24) of ordinary differential equations. We have deliberately chosen this presentation to render the passage to Moyal dynamics more transparent. As a word of caution we stress the double role played by the position and momentum in this approach. They are simultaneously the variables with respect to which one performs the partial derivatives in say eq.(14) as well as the initial conditions (7). 

\vspace{0.3 cm}
\noindent
Let us now turn to Moyal dynamics (6,7). It is well known that

\vspace{0.3 cm}
\noindent
{\underline{\bf Theorem 2.7:}} The formal solution of the Moyal equation is given by:
\begin{equation}
A_M (q,p,t) = U(t) \star A_0 (q,p) \star U (-t), \hspace{1 cm} U(t) \equiv e_{\star}^{i(t-t_0) H (q,p) / \hbar}.
\end{equation}
Here $\phi (\beta , q, p ) \equiv e_{\star}^{\beta B(q,p)}$ is the noncommutative exponential, solution of:
\begin{equation}
\frac{\partial \phi}{\partial \beta} = B \star \phi =\phi \star B, \hspace{1 cm} \phi (\beta =0, q,p) =1, \hspace{0.3 cm} \forall (q,p) \in T^* M \simeq \R^2.
\end{equation}

\vspace{0.3 cm}
\noindent
{\underline{\bf Remark 2.8:}} Contrary to what happens in classical mechanics (cf.eq.(16)), Moyal evolution does not act as a coordinate (local) transformation. In general:
\begin{equation}
A_M (q,p,t) \ne A_0 \left(Q_M (q,p,t), P_M (q,p,t) \right).
\end{equation}
The exception are flows generated by Hamiltonians which are at most polynomials of degree 2 in $(q,p)$. We shall construct an explicit example which corroborates (29) in due course (see remark 4.4 below).

An immediate consequence of theorem 2.7 is the analog of Lemma 2.3:

\vspace{0.3 cm}
\noindent
{\underline{\bf Lemma 2.9:}} The Hamiltonian is conserved under Moyal evolution.

\vspace{0.3 cm}
\noindent
The following, well known result, is the Moyal counterpart of theorem 2.2:

\vspace{0.3 cm}
\noindent
{\underline{\bf Theorem 2.10:}} Moyal evolution 
\begin{equation}
(q,p) \longrightarrow \left(Q= Q_M (q,p,t), P= P_M (q,p,t) \right)
\end{equation}
being a unitary transformation, preserves the Moyal bracket:
\begin{equation}
\left[Q_M (q,p,t), P_M (q,p,t) \right]_{(q,p)} =1.
\end{equation}
Conversely, all solutions of the constraint (31) are unitary transformations \cite{Bracken1}. 
\\
\\
Theorem 2.10 states that it is the Moyal bracket rather than the Poisson bracket that is preserved under time evolution. Let us clarify this point. There are, obviously, transformations which are both canonical and unitary.

\vspace{0.3 cm}
\noindent
{\underline{\bf Example 2.11:}} A transformation which is both canonical and unitary. Let us consider the symplectic transformation:
\begin{equation}
Q(q,p) = a q + bp , \hspace{0.5 cm} P (q,p) = cq + dp,
\end{equation}
where:
\begin{equation}
\det \left(
\begin{array}{c c}
a & b\\
c & d
\end{array}
\right) =1.
\end{equation}
This transformation is both canonical as well as unitary:
\begin{equation}
\left[Q,P \right]_{(q,p)} = \left\{Q,P \right\}_{(q,p)} = 1.
\end{equation}
However, there are transformations which are unitary but not canonical and {\it vice-versa}.

\vspace{0.3 cm}
\noindent
{\underline{\bf Example 2.12:}} A Transformation which is unitary but not canonical. Let us define the transformation:
\begin{equation}
Q(q,p) = \beta e^{q / \beta} , \hspace{1 cm} P (q,p) = e^{- q / \beta} \left[p + \gamma \sinh \left( \frac{2 \beta \pi p}{\hbar} \right) \right],
\end{equation}
where $\beta$, $\gamma$ are positive real constants with dimensions of length and momentum, respectively. This transformation is invertible by the inverse function theorem. The Jacobian reads:
\begin{equation}
\frac{\partial (Q,P)}{\partial (q,p)} = \left\{Q,P \right\}_{(q,p)} = 1 + \frac{ 2 \beta \gamma \pi}{\hbar} \cosh \left( \frac{2 \beta \pi p}{\hbar} \right) >1.
\end{equation}
We conclude that this is not a canonical transformation. On the other hand we have:
\begin{equation}
Q(q,p) \left( \frac{ {\buildrel { \leftarrow}\over\partial}}{\partial q} \frac{ {\buildrel { \rightarrow}\over\partial}}{\partial p} -  \frac{{\buildrel {\leftarrow}\over\partial}}{\partial p}  \frac{{\buildrel { \rightarrow}\over\partial}}{\partial q}  \right)^{2n+1} P (q,p) = \beta \gamma \left( \frac{2 \pi}{\hbar} \right)^{2n+1} \cosh \left( \frac{2 \beta \pi p}{\hbar} \right), \hspace{0.3 cm} n \ge 1,
\end{equation}
where the derivatives ${\buildrel { \leftarrow}\over\partial}$ and ${\buildrel { \rightarrow}\over\partial}$ act on $Q$ and $P$, respectively. 
Consequently:
\begin{equation}
\begin{array}{c}
\left[Q , P \right]_{(q,p)} = \left\{Q , P \right\}_{(q,p)} + \sum_{n=1}^{\infty} \frac{(-1)^n}{(2n+1)!} \left( \frac{\hbar}{2} \right)^{2n} Q(q,p) \left( \frac{ {\buildrel { \leftarrow}\over\partial}}{\partial q} \frac{ {\buildrel { \rightarrow}\over\partial}}{\partial p} -  \frac{{\buildrel {\leftarrow}\over\partial}}{\partial p}  \frac{{\buildrel { \rightarrow}\over\partial}}{\partial q}  \right)^{2n+1} P (q,p) =  \\
\\
= 1 + \frac{ 2 \beta \gamma}{\hbar} \cosh \left( \frac{2 \beta \pi p}{\hbar} \right) \sin \pi =1.
\end{array}
\end{equation}
And so this is a unitary transformation.

As the latter example shows, unitary transformations are not canonical in general. Likewise, canonical transformations are not unitary in general (see example 1 below). Therefore, one should expect Moyal and classical trajectories to diverge. Indeed, this is stated in the following theorem.

\vspace{0.3 cm}
\noindent
{\underline{\bf Theorem 2.13:}} At an arbitrary instant $t \ne t_0$ we have, in general, for $Q= Q_M$ , $P=P_M$:
\begin{equation}
\dot Q (t) \ne  \left[Q (t) , H (Q,P) \right]_{(Q,P)} = \frac{\partial }{\partial P} H (Q,P), \hspace{1 cm} \dot P (t) \ne  \left[P (t) , H (Q,P) \right]_{(Q,P)} = - \frac{\partial }{\partial Q} H (Q,P)  
\end{equation}
   
\vspace{0.3 cm}
\noindent
{\underline{\bf Proof:}} For $A_0(q,p)=q$ the solution of eq.(5) takes the form (cf.(27)) $Q (q,p,t) = U(t) \star q \star U(-t)$. And so:
\begin{equation}
\dot Q (q,p,t) = U(t) \star \left[q, H(q,p) \right]_{(q,p)} \star U(-t) = U(t) \star F_q(q,p) \star U(-t),
\end{equation}
where $F_q(q,p) =\partial H / \partial p$. Unless $F_q(q,p)$ is linear, we have (cf. remark 2.8):
\begin{equation}
U(t) \star F_q(q,p) \star U(-t) \ne F_q \left( Q (q,p,t), P (q,p,t) \right).
\end{equation}
Thus, in general:
\begin{equation}
\dot Q (t) \ne  \frac{\partial }{\partial P} H (Q,P).
\end{equation}
A similar argument holds for $P_M (q,p,t)$.$_{\Box}$

\section{Other approaches to Moyal dynamics}

The main conclusion of the previous section is that the partial differential equations (6) governing Moyal dynamics are not equivalent to the system of ordinary differential equations (9). Since eqs.(6) are of arbitrary order (possibly infinite) the problem of solving them is notoriously difficult. In this section we investigate various alternative formulations of Moyal mechanics. One obvious approach is that of evaluating semiclassical corrections (order by order in $\hbar$) to the classical solutions. In section 3.1 Moyal dynamics is written in terms of a $\hbar$-hierarchy of recursive first order linear partial differential equations. This hierarchy has been previously presented in \cite{Osborn}. Here we re-derive it in other (possibly simpler) terms. In section 3.2 Moyal equations are re-written as a system of ordinary differential equations in the space of noncommutative phase-space functions. This is an interesting formulation allowing, in some cases, to solve Moyal equations exactly (see section 4). Furthermore, and as a by product, Moyal dynamics will be written in terms of a $\hbar$-hierarchy of systems of ordinary first order linear and inhomogeneous differential equations in section 3.3 and 3.4. In a first step this result is derived for analitic Hamiltonians (section 3.3) and then generalized to the case of smooth, locally integrable ones (section 3.4). In this section we shall write $Q_M =Q$, $P_M =P$ for simplicity.

\subsection{A hierarchy of first order partial differential equations}

If we regard Moyal evolution as a partial differential equation, then we can easily derive an equivalent infinite hierarchy of linear inhomogeneous partial differential equations. Indeed let $Q(q,p,t)$ and $P(q,p,t)$ be the solutions of (6,7). A key feature of these solutions is that they display a regular behavior on $\hbar$ and admit a formal asymptotic series:
\begin{equation} 
Q(q,p,t) = \sum_{n=0}^{+ \infty} \hbar^n Q^{(n)} (q,p,t), \hspace{1 cm} P(q,p,t) = \sum_{n=0}^{+ \infty} \hbar^n P^{(n)} (q,p,t),
\end{equation}
where, obviously, $Q^{(0)}(q,p,t)$, $P^{(0)} (q,p,t)$ are the classical solutions. A simple analysis reveals that only the even order terms are non-vanishing, provided the Hamiltonian is independent of $\hbar$. If we substitute (43) in the Moyal equations, we get the classical Hamilton equations for $n=0$ and:
\begin{equation}
\left\{
\begin{array}{l}
\dot Q^{(2n)} = \sum_{k=0}^n \left[Q^{(2k)} , H \right]_{2(n-k)} = \left\{Q^{(2n)} , H \right\}+ \sum_{k=0}^{n-1} \left[Q^{(2k)} , H \right]_{2(n-k)},\quad n\ge 1\\
\\
\dot P^{(2n)} = \sum_{k=0}^n \left[P^{(2k)} , H \right]_{2(n-k)} = \left\{P^{(2n)} , H \right\}+ \sum_{k=0}^{n-1} \left[P^{(2k)} , H \right]_{2(n-k)}, \quad n\ge 1
\end{array}
\right.
\end{equation}
where we used the notation:
\begin{equation}
\left[A(q,p) , B (q,p) \right]_{2n} = \left. \frac{(-1)^n}{(2n+1)! 4^n} \left(\frac{\partial}{\partial q}\frac{\partial}{\partial p'}- \frac{\partial}{\partial p}\frac{\partial}{\partial q'} \right)^{2n +1} A(q,p) B(q',p') \right|_{(q',p') = (q,p)}.
\end{equation}
Here $Q^{(2n)} (t)$ and $P^{(2n)} (t)$ satisfy the initial conditions:
\begin{equation}
Q^{(2n)} (q,p,t=t_0) = P^{(2n)} (q,p,t=t_0)=0, \hspace{0.5 cm} \forall n \ge 1, \hspace{0.2 cm} \forall (q,p) \in \R^2.
\end{equation}
The important thing to remark in eqs.(44) is the fact that they constitute an infinite hierarchy of linear inhomogeneous partial differential equations. For each order $2n$ they are in fact a system of two independent Hamiltonian transport equations with inhomogeneous terms:
\begin{equation}
F_q^{(2n)}(q,p,t)=\sum_{k=0}^{n-1}[Q^{(2k)},H]_{2(n-k)} \quad \mbox{and} \quad 
F_p^{(2n)}(q,p,t)=\sum_{k=0}^{n-1}[P^{(2k)},H]_{2(n-k)}
\end{equation} 
exclusively dependent on the solutions of order $0,2, \cdots, 2n-2$. The solution of eq.(47) is then \cite{Osborn}:
\begin{equation}
\left\{
\begin{array}{l}
Q^{(2n)}(q,p,t)=\int_{t_0}^t d\tau \, F_q^{(2n)}(Q^{(0)}(q,p,\tau),P^{(0)}(q,p,\tau),\tau),\\
\\ 
P^{(2n)}(q,p,t)=\int_{t_0}^t d\tau \, F_p^{(2n)}(Q^{(0)}(q,p,\tau),P^{(0)}(q,p,\tau),\tau).
\end{array}
\right.
\end{equation}

\subsection{Ordinary differential equations in the non-commutative phase space}
 
We start by introducing the concept of $\star$-function. Intuitively, these are functions whose expressions are written in terms of $\star$-products, i.e. the fundamental operations used to express the $\star$-functions are summations, multiplication by scalars and $\star$-products. The most interesting feature of these functions is that their functional form is preserved through Moyal time evolution. Hence they provide the right objects to cast eqs.(8) in an invariant form.  

The main question is then what kind of functions $f:T^*M \longrightarrow \bkC $ can be written as $\star$-functions. 
To begin with the simplest case, let us consider an analytic function $N(q,p)$ in the phase space variables. It is trivial to check that \cite{Dias4} (let $O_1=q$ and $O_2=p$):
\begin{equation}
N(q,p)=\sum_{n=0}^{\infty} \frac{1}{n!} \sum_{1\le i_1,...,i_n \le 2} \left. \frac{\partial^n N}{\partial O_{i_1}...\partial O_{i_n}} \right|_{(q,p)=(0,0)} \left(O_{i_1} \star...\star O_{i_n} \right)_S \equiv [N(q,p)]_S
\end{equation}
where $S$ stands for total symmetrization. We used the fact that $(\underbrace{q\star ...\star q}_{n\, \mbox{terms}} \star \underbrace{p \star ... \star p}_{m\, \mbox{terms}})_S=q^n p^m$ and introduced the notation $[N(q,p)]_S$ to denote the $\star$-function associated to $N(q,p)$ which is written in a completely symmetric form. 
Notice that the objects $[N(q,p)]_S$ are naturally characterized as elements of the enveloping algebra ${\cal U}_{\star} \left( {\cal A} \right)$ of the Heisenberg-Weyl Lie algebra, i.e. as series in the non-commutative coordinates.

Let us now extend the concept of $\star$-function beyond the case of phase-space analytic functions.  We start by noticing that an arbitrary locally integrable function $F(q,p)$ admits a (generalized) Fourier representation: 
\begin{equation}
F(q,p) = \int d \xi \int d \eta \hspace{0.2 cm} \alpha ( \xi, \eta) e^{i \xi q + i \eta p}.
\end{equation}
in terms of possibly distributional coefficients $\alpha ( \xi, \eta)$. 
The set of functions of the form (50) comprises the set of all phase space analytic functions as well as the set of square integrable functions (in fact formula (50) provides a representation for generalized functions). 

All functions of the form (50) can now be written as $\star$-functions:  
\begin{equation}
\left[F(q,p) \right]_S = \int d \xi \int d \eta \hspace{0.2 cm} \alpha ( \xi, \eta) e_{\star}^{i \xi q + i \eta p}.
\end{equation}
For example, if $F(q,p) = q^2p$, then we have $\alpha ( \xi, \eta)= -i \delta''(\xi) \delta' ( \eta)$. It then follows that $\left[F(q,p) \right]_S =  \left( q \star q \star p + q \star p \star q + p \star q \star q \right)/3$.
Notice that the identity $F(q,p)=[F(q,p)]_S$ follows immediately from: 
\begin{equation}
e_{\star}^{i \xi q + i \eta p}= e_{\star}^{i \xi q }\star e_{\star}^{i \eta p} e^{\frac{i \hbar}{2} \xi \eta} = e^{i \xi q + i \eta p}.
\end{equation}
where we used the Baker-Campbell -Hausdorff (BCH) formula.

To proceed we notice that $\star$-functions can be cast in many different functional forms which is a direct consequence of ordering ambiguities (e.g. $q\star p=p\star q +i \hbar$). An important alternative to the complete symmetrization is the standard-antistandard symmetrization that we shall use in section 4 to study a specific example. 
In this case we write a combination of a term, where all the $q$'s stand to the left and its conjugate. For $F(q,p)=q^2 p$, we have $\left[F(q,p) \right]_{SAS}  = \left(q \star q \star p + p \star q \star q \right)/2$. This can be expressed as:
\begin{equation}
\left[F(q,p) \right]_{SAS}  = \frac{1}{2} \int d \xi \int d \eta \hspace{0.2 cm} \alpha (\xi, \eta) \sec \left( \frac{\hbar}{2} \frac{\partial^2}{\partial q \partial p} \right) 
\left( e_{\star}^{i \xi q} \star e_{\star}^{i \eta p} + e_{\star}^{i \eta p} \star e_{\star}^{i \xi q} \right).
\end{equation} 
\\
{\underline{\bf  Lemma 3.1:}} For a phase space function satisfying (50), the following holds:
\begin{equation}
\left[F(q,p) \right]_{SAS} = F(q,p).
\end{equation}

\vspace{0.3 cm}
\noindent
{\underline{\bf Proof:}} Again using the BCH formula, we have:
\begin{equation}
\begin{array}{c}
\frac{1}{2} \int d \xi \int d \eta \hspace{0.2 cm} \alpha (\xi, \eta) \left( e_{\star}^{i \xi q} \star e_{\star}^{i \eta p} + e_{\star}^{i \eta p} \star e_{\star}^{i \xi q} \right) = \frac{1}{2} \int d \xi \int d \eta \hspace{0.2 cm} \alpha (\xi, \eta) e_{\star}^{i \xi q + i \eta p} \left( e^{-i \hbar \xi  \eta /2} + e^{i \hbar \xi  \eta /2}  \right) = \\
\\
= \int d \xi \int d \eta \hspace{0.2 cm} \alpha (\xi, \eta) e^{i \xi q + i \eta p} \cos \left( \frac{\hbar \xi \eta}{2} \right) = \cos \left( \frac{\hbar}{2} \frac{\partial^2}{\partial q \partial p} \right) F(q,p)
\end{array}
\end{equation}
from where the identity (53) follows.$_{\Box}$
\\

Now let us go back to the equations of motion. At time $t=t_0$ eqs.(8) hold. However, they are only valid at $t=t_0$. Let $U (t)$ be the symbol of the evolution operator and let $\left[ F_q (q,p) \right]_{{\cal O}}$ be the $\star$-function associated to $F_q (q,p) = \frac{\partial H}{\partial p}$ and written in the order ${\cal O} =S$ or $SAS$. We then have:
\begin{eqnarray}
\dot Q(t) &=& [Q(t),H]_{(q,p)}=U(t)\star [Q(0),H]_{(q,p)} \star U(-t) \nonumber \\
&=& U (t) \star \left[ F_q (q,p) \right]_{{\cal O}} \star U (-t) =  \left[ F_q \left( Q (t) ,  P (t) \right) \right]_{{\cal O}},
\end{eqnarray}
and a similar equation for the momentum. The important thing to remark is the fact that the equations of motion (56) are now valid for all times in some interval $\left. \right]t_0 - \epsilon, t_0 + \epsilon \left[ \right.$ $(\epsilon >0)$. We may now regard these equations as a system of ordinary differential equations for the position and momentum. We just have to bear in mind that we have to look for solutions in the set of $\star$-functions. The example of section 4 will illustrate this procedure and show how, in some cases, the formulation (56) allows for the exact solution of Moyal dynamics. .

\subsection{A hierarchy of ordinary differential equations for analytic Hamiltonians}

As a direct application of equation (56) we now study the $\hbar$-semiclassical expansion of the Moyal evolution of position and momentum (43). Let us consider the case where the Hamiltonian $H(q,p)$ is a real analytic function, independent of $\hbar$. In this case $[F_q(q,p)]_S$ can be written as:
\begin{equation}
[F_q(q,p)]_S=\sum_{n,m=0}^{\infty} C_{n,m}\left(q^np^m\star^k \right)
\end{equation}
where $k=n+m-1$ for $(n,m)\not=(0,0)$ and $k=0$ for $n=m=0$, $C_{n,m}$ are the Taylor coefficients and:
\begin{equation} 
(q^np^m \star ^k) \equiv (\underbrace{q\star ...\star q}_{n\, \mbox{terms}} \star \underbrace{p \star ... \star p}_{m\, \mbox{terms}})_S.
\end{equation}
Using this notation eq.(56) is written:
\begin{equation} 
\dot Q= \sum_{n,m=0}^{\infty} C_{n,m} (Q^n P^m \star ^k)
\end{equation}
Let us then expand this equation in powers of $\hbar$. We consider the expansions (43) and introduce the additional notation:
\begin{equation}
\star =\sum_{n=0}^{+\infty} \hbar^n \star_n \quad , \quad 
A(q,p)\star_n B(q,p)=\frac{1}{n!} \left(\frac{i}{2} \right)^n \left. \left(\frac{\partial}{\partial q}\frac{\partial}{\partial p'}- \frac{\partial}{\partial p}\frac{\partial}{\partial q'} \right)^n A(q,p) B(q',p') \right|_{(q',p') = (q,p)}.
\end{equation}
It then follows that:
\begin{equation}
(Q^n P^m \star ^k)=\sum_{r=0}^{\infty} \hbar ^r (Q^n P^m \star ^k)_r
\end{equation}
where 
\begin{equation}
(Q^n P^m \star ^k)_r= \sum_{\begin{array}{c}
\sum_{\alpha =1}^n i_{\alpha} +
\sum_{\beta =1}^m j_{\beta} +\sum_{\gamma =1}^k l_{\gamma} =r \\
i_{\alpha},j_{\beta},l_{\gamma} \in \{0,...,r\}
\end{array}}
\left(Q^{(i_1)} \star_{l_1} Q^{(i_2)} \star_{l_2} \cdots \star_{l_{n-1}}Q^{(i_n)} \star_{l_n} P^{(j_1)} \star_{l_{l+1}} \cdots  \star_{l_k} P^{(j_m)} \right)_S  
\end{equation}
and so the leading order eq.(59,62) is the classical Hamiltonian equation, and to order ${\cal O} (\hbar^r)$, $r\ge 1$ it reads:
\begin{eqnarray}
\dot Q^{(r)} &= & \sum_{n,m=0}^{\infty} C_{n,m} (Q^nP^m\star^k)_r \\
&=& 
\sum_{n,m=0}^{\infty} C_{n,m} \left( nQ^{(r)} (Q^{(0)})^{n-1} (P^{(0)})^m  +m (Q^{(0)})^n P^{(r)} (P^{(0)})^{m-1} \right) + f_r(Q^{(0)},...,Q^{(r-1)},P^{(0)},...,P^{(r-1)}) \nonumber
\end{eqnarray}
together with the initial conditions:
\begin{equation}
Q^{(r)} (t=t_0)=P^{(r)} (t=t_0)=0 , \quad r\ge 1
\end{equation}
and the inhomogeneous term is:
\begin{eqnarray}
&& f_r(Q^{(0)},...,Q^{(r-1)},P^{(0)},...,P^{(r-1)}) \\
&=& \sum_{n,m=0}^{\infty} C_{n,m} \sum_{ \begin{array}{c}
\sum_{\alpha =1}^n i_{\alpha} +
\sum_{\beta =1}^m j_{\beta} +\sum_{\gamma =1}^k l_{\gamma} =r \\
i_{\alpha},j_{\beta} \in \{0,...,r-1\} ; l_{\gamma} \in \{0,...r\}
\end{array}}
\left(Q^{(i_1)} \star_{l_1} Q^{(i_2)} \star_{l_2} \cdots \star_{l_{n-1}}Q^{(i_n)} \star_{l_n} P^{(j_1)} \star_{l_{n+1}} \cdots \star_{l_k}P^{(j_m)}\right)_S \nonumber 
\end{eqnarray}
A similar set of equations are valid for $P$. Finally, it is easy to verify that the initial conditions will impose the odd contributions to the semiclassical expansions of $Q,P$ to be identically zero.

\subsection{The hierarchy of ordinary differential equations in the general case}

We now generalize the previous approach to local integrable Hamiltonians (still real and time, $\hbar$ independent). 
 
\vspace{0.3 cm}
\noindent
{\underline{\bf Theorem 3.2:}} Let the Hamiltonian admit a Fourier representation as in (50). Then in the space of formal series $\left[ \right. \left[ \right. \hbar  \left. \right] \left. \right]$ the Moyal evolution of position and momentum is dictated by a hierarchy of sets of inhomogeneous first order linear differential equations.

\vspace{0.3 cm}
\noindent
{\underline{\bf Proof:}} Let us write $z=(q,p)$, $Z=(Q,P)$. The Moyal equations at time $t=t_0$ read:
\begin{equation}
\dot Z_i (t=t_0) = F_i, \hspace{1 cm} F_i = J_{ij} \partial_j H (z),
\end{equation}
where sum over repeated indices is understood, $H$ is the Hamiltonian and we used the compact notation $\partial_i = \left(\frac{\partial}{\partial q}, \frac{\partial}{\partial p} \right)$ , $J_{qq}=J_{pp}=0$, $J_{qp}=-J_{pq}=1$. Under the assumptions of the theorem the vector function $F$ admits the Fourier transform:
\begin{equation}
F(z) = \int d \xi \hspace{0.2 cm} f (\xi) e^{i \xi \cdot z},
\end{equation}
where $\xi = (\xi_q , \xi_p)$ and $\xi \cdot z= q \xi_q + p \xi_p$. Following the procedure discussed in section 3.2, we may write at time $t \ge t_0$:
\begin{equation}
\dot Z (t) = \int d \xi \hspace{0.2 cm} f (\xi) e_{\star}^{i \xi \cdot Z (t)}.
\end{equation}
In ref.\cite{Dias3} we proved that the noncommutative exponential may be expanded in the form:
\begin{equation}
e_{\star}^B = \left( 1 + \hbar^2 A_2 + \hbar^4 A_4 + \cdots \right)  e^B,
\end{equation}
where $A_2,A_4, \cdots$ depend only on $B(z)$ and its derivatives. For instance:
\begin{equation}
A_2 (z) = - J_{ik}J_{jl} \left(\partial_i \partial_j B\right) \left[\frac{1}{16} \left(\partial_k \partial_l B\right) + \frac{1}{24} \left(\partial_k B \right) \left(\partial_l B \right)  \right].
\end{equation}
It should be noted that $B$ may also depend upon $\hbar$. If we expand the Moyal trajectories
\begin{equation}
Z(t) = Z^{(0)} (t)+ \hbar^2 Z^{(2)} (t) +  \hbar^4 Z^{(4)} (t) +\cdots
\end{equation}
as well as the coefficients of the exponential
\begin{equation}
A_{2k} (Z) = A_{2k}^{(0)} (Z^{(0)}) + \hbar^2 A_{2k}^{(2)} (Z^{(0)}, Z^{(2)})  + \hbar^4 A_{2k}^{(4)} (Z^{(0)}, Z^{(2)}, Z^{(4)} ) + \cdots
\quad, k=1,2,...
\end{equation}
we get from (68):
\begin{equation}
\begin{array}{c}
\dot Z^{(0)} (t)+ \hbar^2 \dot Z^{(2)} (t) +  \hbar^4 \dot Z^{(4)} (t)  +  \hbar^6 \dot Z^{(6)} (t)+\cdots = \int d \xi \hspace{0.2 cm} f(\xi) \left\{e^{i \xi \cdot \left( Z^{(0)} + \hbar^2 Z^{(2)}  +  \hbar^4 Z^{(4)} +  \hbar^6 Z^{(6)}  +\cdots \right)} + \right.\\
\\
 +\hbar^2 \left[A_2^{(0)} (Z^{(0)}) + \hbar^2 A_2^{(2)}(Z^{(0)}, Z^{(2)}) + \hbar^4 A_2^{(4)} (Z^{(0)}, Z^{(2)}, Z^{(4)} ) + \cdots \right] e^{i \xi \cdot \left( Z^{(0)} + \hbar^2 Z^{(2)} + \hbar^2 Z^{(4)} +\cdots \right)} + \\
\\
\left. + \hbar^4 \left[A_4^{(0)} (Z^{(0)} ) +\hbar^2 A_4^{(2)} (Z^{(0)}, Z^{(2)} )  + \cdots \right] e^{i \xi \cdot \left( Z^{(0)} + \hbar^2 Z^{(2)} +\cdots \right)} + \hbar^6 \left[A_6^{(0)} (Z^{(0)}) + \cdots \right] e^{i \xi \cdot \left( Z^{(0)} + \cdots \right)} + \cdots \right\}.
\end{array}
\end{equation}
And so to order $2n$:
\begin{equation}
\dot Z_r^{(2n)} (t) = \int d \xi \hspace{0.2 cm} f_r(\xi) e^{i \xi \cdot Z^{(0)}} \left(i \xi \cdot Z^{(2n)} + \cdots \right) = Z_i^{(2n)} (t) \cdot \frac{\partial F_r }{\partial Z_i^{(0)}} (Z^{(0) } (t))  + \cdots    
\end{equation}
where the dots stand for terms which depend only on $Z^{(0)}, Z^{(2)}, \cdots, Z^{(2n-2)}$. As advertised these constitute an infinite hierarchy of ordinary first order linear differential equations.$_{\Box}$

\vspace{0.3 cm}
\noindent
{\underline{\bf Remark 3.3:}} The theorem is stated in the space of formal series, which means that questions of convergence are ignored (this was already the case in sections 3.1 and 3.3). The important thing to stress is the fact that the hierarchy of {\it partial} differential equations of section 3.1 has now been transformed into a hierarchy of {\it ordinary} differential equations. The proof also indicates how one can write the inhomogeneous term in the system of ordinary differential equations to each order ${\cal O} (\hbar^{2n} )$. For instance, to order ${\cal O} (\hbar^2 )$ the inhomogeneous term reads:
\begin{equation}
\begin{array}{c}
\int d \xi \hspace{0.2 cm} f (\xi) e^{i \xi \cdot Z^{(0)}} A_2^{(0)}(Z^{(0)} ) = \\
\\
= -\int d \xi \hspace{0.2 cm} f (\xi) e^{i \xi \cdot Z^{(0)}} J_{ik}J_{jl}\left[\partial_i \partial_j \left(i \xi \cdot Z^{(0)} \right) \right] \times \left[\frac{1}{16} \partial_k \partial_l \left(i \xi \cdot Z^{(0)} \right) + \frac{1}{24} \partial_k  \left(i \xi \cdot Z^{(0)} \right) \partial_l \left(i \xi \cdot Z^{(0)} \right) \right] = \\
\\
= - J_{ik} J_{jl} \left(\partial_i \partial_j Z_a^{(0)} \right) \times \left[ \frac{1}{16} \left( \partial_k \partial_l Z_b^{(0)}\right) \frac{\partial}{\partial Z_b^{(0)}} + \frac{1}{24} \left( \partial_k Z_b^{(0)}\right) \left(\partial_l Z_c^{(0)}\right)  \frac{\partial^2}{\partial Z_b^{(0)} Z_c^{(0)}} \right] \frac{\partial}{\partial Z_a^{(0)}}   \int d \xi \hspace{0.2 cm} f (\xi) e^{i \xi \cdot Z^{(0)}}.
\end{array}
\end{equation}
Altogether we may then write:
\begin{equation}
\begin{array}{c}
\dot Z_r^{(2)} (t) = Z_i^{(2)} (t) \frac{\partial F_r}{\partial Z_i^{(0)}}(Z^{(0)}) - \frac{1}{16} J_{ik} J_{jl} \left( \partial_i \partial_j Z_a^{(0)} \right) \left( \partial_k \partial_l Z_b^{(0)} \right) \frac{\partial^2 F_r}{\partial Z_a^{(0)} \partial Z_b^{(0)}} (Z^{(0)} ) \\
\\
-  \frac{1}{24} J_{ik} J_{jl} \left( \partial_i \partial_j Z_a^{(0)} \right) \left( \partial_k  Z_b^{(0)} \right) \left(\partial_l Z_c^{(0)} \right) \frac{\partial^3 F_r}{\partial Z_a^{(0)} \partial Z_b^{(0)} \partial Z_c^{(0)}} (Z^{(0)} ).
\end{array}
\end{equation}
This equation could be equally derived using the method of section 3.3 for analytic Hamiltonians. 

\section{Example 1}

The following example shows explicitly that Moyal and classical evolutions are different. Moreover it will illustrate the formalism developed in section 3.2. We shall assume $t_0=0$ for simplicity and we start by computing the classical evolution generated by the following Hamiltonian with position dependent mass:
\begin{equation}
H(q,p) = \frac{q^2 p^2}{4 m l^2},
\end{equation}
where $l$ is some real constant with dimensions of length. The Hamilton equations read:
\begin{equation}
\dot Q_C = \frac{Q_C^2 P_C}{2 m l^2}, \hspace{1 cm} \dot P_C = - \frac{Q_C P_C^2}{2 m l^2}.
\end{equation}
We conclude that $Q_C \cdot P_C$ is a constant of motion:
\begin{equation}
Q_C (q,p,t) \cdot P_C (q,p,t) = q \cdot p , \hspace{0.5 cm} \forall t.
\end{equation}
Substituting this expression in the first equation (78) we get:
\begin{equation}
\dot Q_C (t) = \frac{qp}{2 m l^2} Q_C (t).
\end{equation}
And so, from (7):
\begin{equation}
Q_C (q,p,t) = q e^{\frac{qpt}{2 m l^2}}, \hspace{0.5 cm} \forall t \in \R.
\end{equation}
Upon substitution in (79), we get:
\begin{equation}
P_C (q,p,t) = p e^{-\frac{qpt}{2 m l^2}}, \hspace{0.5 cm} \forall t \in \R.
\end{equation}
By construction this is a canonical transformation at all times. However, this is not a unitary transformation. To check this, let us first compute $Q_C \star P_C$ using the kernel representation (2). If we Wick rotate $t \to i \tau$, we get after some integrations and an integration by parts:
\begin{equation}
Q_C(q,p,t) \star_{(q,p)} P_C (q,p,t) = \frac{qp + i \hbar/2}{\left[1 + \left( \frac{ \hbar t}{4 m l^2} \right)^2 \right]^2} =  \frac{q \star_{(q,p)}p }{\left[1 + \left( \frac{ \hbar t}{4 m l^2} \right)^2 \right]^2} .
\end{equation}
It then follows that:
\begin{equation}
\left[Q_C(q,p,t), P_C (q,p,t) \right]_{(q,p)} = \frac{2}{ \hbar} Im \left(Q \star_{(q,p)} P \right) = \left[1 + \left( \frac{ \hbar t}{4 m l^2} \right)^2 \right]^{-2} \ne 1, \hspace{0.5 cm} \forall t \ne 0.
\end{equation}
And so this is not a unitary transformation. From (31) we conclude that the Moyal evolution must differ from the classical solution. The Moyal evolution for an arbitrary observable $A$ is dictated by the equation:
\begin{equation}
\dot A_M = \left[\frac{qp}{m l^2} \left(q \frac{\partial }{\partial q}- p \frac{\partial }{\partial p} \right)  + \frac{ \hbar^2}{8 m l^2} \left( q \frac{\partial^3 }{\partial q^2 \partial p} - p \frac{\partial^3 }{\partial q \partial p^2}   \right) \right] A_M.
\end{equation}
Now this can be solved for $\left(Q_M, P_M \right)$ with the initial conditions (7). This is feasible in this case. But, in general, it is the fact that we are dealing with a partial differential equation that makes Moyal evolution difficult to solve. We shall follow the alternative route presented in section 3.2. This approach is reminiscent of the classical solution presented above. We shall then solve an ordinary differential equation in the noncommutative space of $\star$-functions. The equations of motion at time $t=0$ read (cf.(8)):
\begin{equation}
\dot Q = \frac{q^2 p}{2 m l^2}, \hspace{1 cm} \dot P = - \frac{q p^2}{2 m l^2}.
\end{equation}
These equations are valid only at time $t=0$. To write the corresponding ordinary differential equations at another time $t$ we will have to write the right-hand sides in terms of $\star$-products. We resort to lemma 3.1 and write the equations (86) in the SAS order:
\begin{equation}
\dot Q = \frac{1}{4 m l^2} \left(q \star q \star p + p \star q \star q \right) , \hspace{1 cm}
\dot P = - \frac{1}{4 m l^2} \left(q \star p \star p + p \star p \star q \right).
\end{equation}
If $U(t)$ is as in (27), then if we act with $U(t) \star$ on the left and with $\star U(-t)$ on the right of these equations, we get:
\begin{equation}
\dot Q_M = \frac{1}{4 m l^2} \left(Q_M \star Q_M \star P_M + P_M \star Q_M \star Q_M \right) , \hspace{1 cm}
\dot P_M = - \frac{1}{4 m l^2} \left(Q_M \star P_M \star P_M + P_M \star P_M \star Q_M \right).
\end{equation}
The important thing to remark is that, contrary to eqs.(87), the previous expressions hold for  all $t \in \left. \right] - \epsilon , \epsilon \left[ \right.$ with $\epsilon >0$\footnote{We show below that $\epsilon = \frac{2 \pi m l^2}{\hbar}$}. We may thus regard eqs.(88) as ordinary differential equations for $\left(Q_M, P_M \right)$. We just have to bear in mind the fact that we have to find solutions in the space of $\star$-functions . Since $\left[Q_M (t), P_M (t) \right]=1$ at any time $t$, we have:
\begin{equation}
\begin{array}{c}
\frac{d}{dt} \left(Q_M \star P_M \right) = \dot Q_M \star P_M + Q_M \star \dot P_M = \frac{1}{4 m l^2} \left(P_M \star Q_M \star Q_M \star P_M - Q_M \star P_M \star P_M \star Q_M \right)=\\
\\
= \frac{1}{4 m l^2} \left(P_M \star Q_M \star Q_M \star P_M - P_M \star Q_M \star Q_M \star P_M + i \hbar P_M \star Q_M - i \hbar P_M \star Q_M  \right)=0.
\end{array}
\end{equation}
And so, as in the classical case, $Q_M \star P_M$ is a constant of motion:
\begin{equation}
Q_M (q,p,t) \star_{(q,p)} P_M (q,p,t) =q \star p = qp + \frac{i \hbar}{2}.
\end{equation}
Substituting (90) in the first eq.(88):
\begin{equation}
\dot Q_M (t) = \frac{1}{4 m l^2} \left[Q_M (t) \star (qp) + (qp) \star Q_M (t) \right].
\end{equation}
The solution of this equation subject to the initial condition (7) is:
\begin{equation}
Q_M (q,p,t) = e_{\star}^{\frac{tqp}{4 m l^2}} \star q \star e_{\star}^{\frac{tqp}{4 m l^2}}.
\end{equation}
Similarly, we get for $P_M$:
\begin{equation}
P_M (q,p,t) = e_{\star}^{-\frac{tqp}{4 m l^2}} \star p \star e_{\star}^{-\frac{tqp}{4 m l^2}}.
\end{equation}
By construction this is a unitary transformation. To finish our calculation, we just need to compute explicitly the $\star$-products in (92,93). The noncomutative exponentials have been computed for quadratic functions in \cite{Dias3}, \cite{Flato}. The result is:
\begin{equation}
\phi_{\pm} (t,q,p) \equiv e_{\star}^{\pm \frac{tqp}{4 m l^2}} = \sec^2 \left(\frac{\hbar t}{8 m l^2} \right) \exp \left[\pm  \frac{2}{\hbar} qp \tan \left(\frac{\hbar t}{8 m l^2} \right) \right], \hspace{1 cm} |t| < \frac{4 \pi m l^2}{\hbar}.
\end{equation} 
And so:
\begin{equation}
\begin{array}{c}
Q_M (q,p,t) = \phi_+ \star q \star \phi_+ = \left[1 + i \tan \left(\frac{\hbar t}{8 m l^2} \right) \right] \phi_+ \star \left(q \phi_+ \right) = \\
\\
= \sec^2 \left(\frac{\hbar t}{4 m l^2} \right) q \exp \left[ \frac{2}{\hbar} qp \tan \left(\frac{\hbar t}{4 m l^2} \right) \right], \hspace{1 cm} |t| < \frac{2 \pi m l^2}{\hbar},
\end{array}
\end{equation} 
where the last $\star$-product was again evaluated using the kernel representation (2). A similar calculation leads to:
\begin{equation}
P_M (q,p,t) = \sec^2 \left(\frac{\hbar t}{4 m l^2} \right) p \exp \left[- \frac{2}{\hbar} qp \tan \left(\frac{\hbar t}{4 m l^2} \right) \right], \hspace{1 cm} |t| < \frac{2 \pi m l^2}{\hbar},
\end{equation} 

\vspace{0.3 cm}
\noindent
{\underline{\bf Remark 4.1:}} The mapping $(q,p) \rightarrow \left( Q_M (t), P_M (t) \right)$ for this system constitutes another example of a transformation which is unitary but not canonical. Indeed we have:
\begin{equation}
\left\{Q_M (q,p,t), P_M (q,p,t) \right\}_{(q,p)} = \sec^4 \left( \frac{\hbar t}{4 m l^2} \right) \ne 1, \hspace{0.5 cm} \forall t \in \left. \right]- \frac{2 \pi m l^2}{\hbar}, \frac{2 \pi m l^2}{\hbar} \left[ \right. \backslash \left\{0 \right\}.
\end{equation}

\vspace{0.3 cm}
\noindent
{\underline{\bf Remark 4.2:}} As expected in the limit $\hbar \to 0$, the previous solutions coincide with the classical ones $\left(Q_C (q,p,t) , P_C (q,p,t) \right)$ (cf.(81,82)). Moreover, we have checked explicitly that the solutions to order $\hbar^2$ $\left(Q^{(2)} (t), P^{(2)} (t) \right)$ satisfy the ordinary differential equations (76).

\vspace{0.3 cm}
\noindent
{\underline{\bf Remark 4.3:}} Finding solutions in the noncommutative space of $\star$-functions might be more restrictive than in classical mechanics. In our example, the Moyal solutions are only valid on the bounded interval $t \in \left. \right]- \frac{2 \pi m l^2}{\hbar}, \frac{2 \pi m l^2}{\hbar} \left[ \right.$, whereas the classical solutions are valid everywhere on $\R$. 
This is a very important point. It highlights the danger of relying too heavily upon $\hbar$-expansions. 
Indeed, if we consider say the second order solution for the position,
$$
Q_M^{(2)} (q,p,t)  = Q_C (q,p,t) \frac{ t^2}{16m^2 l^4} \left(1+ \frac{tqp}{6m l^2} \right),
$$
we realize that it is a smooth function everywhere on $T^* M \simeq \R^2$ and for all $t \in \R$. The same can be said about all the remaining orders. However, our exact solution (95) is singular at $|t| = \frac{2 \pi m l^2}{\hbar}$ in sharp contrast with the perturbative result.

\vspace{0.3 cm}
\noindent
{\underline{\bf Remark 4.4:}} This example illustrates the claim that Moyal evolution does not act as a coordinate (local) transformation. Let us consider the phase space variable:
\begin{equation}
A_0 (q,p) = q \star p = qp + \frac{i \hbar}{2}.
\end{equation}
Since this is a constant of motion, we have:
\begin{equation}
A_M (q,p,t) = U(t) \star A_0 \star U(-t) = Q_M (t) \star P_M (t) = qp + \frac{i \hbar}{2}.
\end{equation}
Inverting (95,96) we get:
\begin{equation}
\left\{
\begin{array}{l}
q \left( Q_M , P_M, t \right) = \cos^2 \left(\frac{\hbar t}{ 4 m l^2} \right) Q_M \exp \left[ - \frac{1}{\hbar} Q_M P_M \sin \left(\frac{\hbar t}{ 2 m l^2} \right) \cos^2 \left(\frac{\hbar t}{ 4 m l^2} \right)\right]\\
\\
p \left( Q_M , P_M, t \right) = \cos^2 \left(\frac{\hbar t}{ 4 m l^2} \right) P_M \exp \left[  \frac{1}{\hbar} Q_M P_M  \sin \left(\frac{\hbar t}{ 2 m l^2} \right) \cos^2 \left(\frac{\hbar t}{ 4 m l^2} \right) \right]
\end{array}
\right.
\end{equation}
Upon substitution in (99) we obtain:
\begin{equation}
A_M = \cos^4 \left(\frac{\hbar t}{ 4 m l^2} \right)  Q_M P_M +  \frac{i \hbar}{2}.
\end{equation}
If Moyal evolution were a coordinate transformation we ought to have (cf(16)):
\begin{equation}
A_M (t) = A_0 \left( Q_M (t), P_M (t) \right) = Q_M P_M + \frac{i \hbar}{2}.
\end{equation}
Notice also that eq.(100) is not the time reversal of eqs.(95,96) (i.e. cannot be obtained by making $t \to -t$) as it would be the case if Moyal evolution were a coordinate transformation.

\section{Example 2}
 
The previous example is admittedly somewhat contrived. It has the virtue of being exactly solvable and of yielding distinct classical and Moyal evolutions. However, the claim that these evolutions are different in general remains valid even if we consider more realistic systems. For Hamiltonians which are polynomial of $q$'s and $p$'s of degree higher than 2, there will generally appear discrepancies between the classical and Moyal evolutions of order ${\cal O} (\hbar^2)$. Indeed, let us consider a Hamiltonian of the form:
\begin{equation}
H (q,p) = \frac{p^2}{2m} + V(q),
\end{equation}
and define ($z=(q,p)$):
\begin{equation}
\left\{
\begin{array}{l}
\Omega_1^z (z) \equiv \left[z, H(z) \right]_z, \hspace{0.5 cm} \Omega_2^z (q,p) \equiv \left[ \Omega_1^z (z) , H (z) \right]_z , \cdots \\
\\
\Lambda_1^z (z) \equiv \left\{z, H(z) \right\}_z, \hspace{0.5 cm} \Lambda_2^z (z) \equiv \left\{ \Lambda_1^z (z), H (z) \right\}_z , \cdots
\end{array}
\right.
\end{equation}
A simple calculation reveals that:
\begin{equation}
\Omega_i^q (z) = \Lambda_i^q (z), \hspace{0.2 cm} i= 1, \cdots, 5, \hspace{1 cm} \Omega_6^q (z) = \Lambda_6^q (z) - \frac{\hbar^2}{4 m^4} V^{(3)} (q) V^{(4)} (q)
\end{equation}
where $V^{(n)} (q)$ is the $n$-th derivative of $V(q)$. Likewise for the momentum we get:
\begin{equation}
\Omega_i^p (z) = \Lambda_i^p (z), \hspace{0.2 cm} i= 1, \cdots, 4, \hspace{1 cm} \Omega_5^p (z) = \Lambda_5^p (z) - \frac{\hbar^2}{4 m^3} V^{(3)} (q) V^{(4)} (q).
\end{equation}
The solutions of the classical $(Z_C (z,t))$ and of the Moyal $(Z_M (z,t))$ equations of motion admit the formal expansions:
\begin{equation}
Z_C (z,t) = z + \sum_{n=1}^{\infty} \frac{t}{n!} \Lambda_n^z (z), \hspace{0.5 cm} Z_M (z,t) = z + \sum_{n=1}^{\infty} \frac{t}{n!} \Omega_n^z (z).
\end{equation}
By virtue of (105-107), we conclude that $Q_M$ diverges from $Q_C$ at order ${\cal O} (t^6)$:
\begin{equation}
Q_M (q,p,t) = Q_C (q,p,t) - \frac{\hbar^2 t^6}{4 \cdot 6! m^4} V^{(3)} (q) V^{(4)} (q) + {\cal O} (t^7),
\end{equation}
and $P_M$ diverges from $P_C$ at order ${\cal O} (t^5)$:
\begin{equation}
P_M (q,p,t) = P_C (q,p,t) - \frac{\hbar^2 t^5}{4 \cdot 5! m^3} V^{(3)} (q) V^{(4)} (q) + {\cal O} (t^6).
\end{equation}
For the quartic anharmonic oscillator $V(q) = \frac{1}{2}m \omega^2 q^2 + \frac{\lambda}{4! } q^4$ $(\lambda >0)$ and thus the previous equations yield: 
\begin{equation}
Q_M (q,p,t) = Q_C (q,p,t) - \frac{\hbar^2 \lambda^2 t^6}{4 \cdot 6!  m^4} q + {\cal O} (t^7), \hspace{1 cm} P_M (q,p,t) = P_C (q,p,t) - \frac{\hbar^2 \lambda^2 t^5}{4 \cdot 5!  m^3} q + {\cal O} (t^6).
\end{equation}
It may appear from this analysis that a cubic potential generates a classical trajectory in Moyal mechanics. However, even this is not true. If we study the momentum to order ${\cal O} (t^7)$ we obtain a term of the form: $\frac{5 \hbar^2 t^7}{4 \cdot 7! m^4} \left[V^{(3)} (q) \right]^3$.

\vspace{0.2 cm}
\noindent
This example illustrates explicitly that the Moyal and classical evolutions of the fundamental variables $(q,p)$ are generally different.

\section{Conclusions and outlook}

We have shown explicitly with various counter examples that, in general, Moyal and classical dynamics differ for the canonical variables of position and momentum. We argued that this stems from the fact that Hamilton's equations are only valid at some initial time for the Moyal evolution. This is a consequence of two related concepts, namely that Moyal evolution is neither a canonical nor a coordinate (local) transformation. A necessary condition for both evolutions to coincide is that Moyal (and classical) evolution be both a canonical and a unitary transformation. Whether this is also sufficient deserves further investigation. Indeed if this condition were sufficient for coincidence of the two evolutions, then in a first instance it would suffice to analyze the classical solution only in order to infer beforehand whether the Moyal evolution is equal to its classical counterpart or not. There are a few comments that can be made in this respect. Let us assume for a while that there exist Moyal trajectories which are simultaneously canonical transformations and distinct from the classical counterparts. Let us consider equations (40):
\begin{equation}
\dot Q = F_q (q,p,t) , \hspace{1 cm} \dot P = F_p (q,p,t),
\end{equation} 
where $F_i (z,t) = U(t) \star \left[z_i,H \right] \star U(-t)$. The solutions are such that $Q= Q (q,p,t)$ and $P= P (q,p,t)$. We may invert these relations and get $q= q(Q,P,t)$, $p= p(Q,P,t)$. Substituting these equations in (111) we get:
\begin{equation} 
\dot Q = G_q (Q,P,t), \hspace{1 cm} \dot P = G_p (Q,P,t),
\end{equation}
where $G_i (Q,P,t) = F_i \left( q (Q,P,t), p (Q,P,t) , t \right)$. In classical mechanics one would expect: $G_q (Q,P,t) = G_q (Q,P,0) = F_q (Q,P,0)= \frac{\partial}{\partial P} H(Q,P)$, etc. If the Moyal evolution is a canonical transformation then there should exist some function $H_E (Q,P)$ (an effective Hamiltonian) such that:
\begin{equation}
G_q (Q,P) = \frac{\partial H_E}{\partial P}, \hspace{1 cm} G_p (Q,P) = - \frac{\partial H_E}{\partial Q}.
\end{equation}
Then the Moyal equations would be equivalent to:
\begin{equation}
\dot Q = \frac{\partial H_E}{\partial P}, \hspace{1 cm} \dot P = - \frac{\partial H_E}{\partial Q}.
\end{equation}
Since at $t=t_0$ we have $H_E (Q,P,t=t_0)= H(q,p)$, we conclude that if $H_E$ is not explicitly time dependent, then, in view of (114), Moyal and classical evolutions would have to coincide. At this point it is not clear to us whether there exist Moyal trajectories which are a canonical transformation and different from the classical trajectories generated by the same Hamiltonian. What can be said from the previous analysis is that if such trajectories do exist then they are equivalently generated by a classical flow with a time dependent Hamiltonian. 

\vspace{0.3 cm}
\noindent
Another byproduct of our analysis is following conjecture: whenever a classical system can be solved exactly in the space of smooth phase space functions with the usual commutative product $\cdot$ and Poisson bracket $\left\{, \right\}$, then the corresponding Moyal flow can be "{\it solved}" in the space of $\star$-functions with noncommutative $\star$-product and Moyal bracket $\left[, \right]$. One just has to deform the classical solution by introducing judiciously $\star$-products (cf.(81,82) and (95,96)). However, it is still unclear what is the precise prescription for substituting standard products by $\star$-products in this context. An even if one finds such prescription the final result in terms of $\star$-functions is obviously not explicit: it is expressed in terms of $\star$-products which may be difficult (or impossible) to evaluate. The obvious question is whether one gains anything by doing so. Indeed, one may just as well express the solution in terms of $\star$-products as in eq.(27). However, by inspection of our example 1, we realize two important facts. First of all in our solution (95,96) one has to evaluate the noncommutative exponential $\phi_{\pm}$ (94) for a quadratic variable $qp$, which is well known \cite{Flato}, \cite{Dias3}, whereas in the universal solution (27) one needs to compute the noncommutative exponential for a quartic variable $H \propto q^2 p^2$ (i.e. the formalism of $\star$-functions considerably simplified the resolution of the problem). Secondly, if one is not capable of evaluating all the $\star$-products in the solution, then our prescription is more apt for performing the $\hbar$-expansions order-by-order. Indeed, for the formal solution (27), when one computes the $\hbar$-expansion, one has to bear in mind the fact that the Weyl symbol of the evolution operator $U(t)$ may be singular in $\hbar$.

Another fact that our example 1 illustrates is that the classical and Moyal trajectories may diverge dramatically during a finite lapse of time and that the perturbative $\hbar$-expansions tend to obscure the global behavior of the trajectories. If we consider the ordinary linear differential equations (63,74) derived in section 3,  the usual existence and uniqueness theorems apply. Typically the range of validity of the solutions found at each order will be less restrictive than that of the global solution. This is obviously connected with questions of convergence of the series, when one re-sums all the terms. We feel that the suggested method of solving the equations first in the noncommutative algebra ${\cal U}_{\star} \left({\cal A} \right)$ may be a better starting point for analyzing the global aspects of Moyal trajectories.
 
\vspace{1 cm}

\begin{center}

{\large{{\bf Acknowledgments}}} 

\end{center}

\vspace{0.3 cm}
\noindent
This work was partially supported by the grants POCTI/MAT/45306/2002 and POCTI/0208/2003 of the Portuguese Science Foundation.

\vspace{1 cm}

\begin{center}

{\large{{\bf Note added}}} 

\end{center}

\vspace{0.3 cm}
\noindent
After this work was completed we have learned about the paper \cite{Krivoruchenko}. Several of the results of that paper are coincident with ours, although the two approaches seem to be complementary. In addition, we present several examples.

\end{document}